\documentclass[aps,prb,reprint,showpacs,longbibliography,superscriptaddress]{revtex4-2}
\usepackage{bm}
\usepackage{graphicx}
\usepackage{epstopdf}
\usepackage{latexsym}
\usepackage{subfigure}
\usepackage{color}
\usepackage{hyperref}
\usepackage{amssymb}
\usepackage{amsmath}
\usepackage{amsbsy,bm}
\usepackage{fancyhdr}
\usepackage[T1]{fontenc}
\newcommand{\CsCoBr}{{$\rm Cs_2CoBr_4$\ }}
\begin{document}

\title{Spin dynamics in ordered phases of anisotropic triangular-lattice antiferromagnet Cs$_{2}$CoBr$_{4}$}

\author{T.~A.~Soldatov}
\affiliation{P.~L.~Kapitza Institute for Physical Problems RAS, 119334 Moscow, Russia}

\author{A.~I.~Smirnov}
\affiliation{P.~L.~Kapitza Institute for Physical Problems RAS, 119334 Moscow, Russia}

\author{A.~V.~Syromyatnikov}
\affiliation{Petersburg Nuclear Physics Institute named by B.P.\ Konstantinov of National Research Center "Kurchatov Institute", Gatchina 188300, Russia}

\begin{abstract}

We study spin dynamics of ordered phases of \CsCoBr in a magnetic field using electron spin resonance (ESR) technique and theoretical analysis. This material hosts weakly
interacting distorted-triangular-lattice planes of spin-$\frac32$ $\rm Co^{2+}$ ions which can be viewed as spin chains coupled by frustrating interactions. Strong single-ion
anisotropy allows to describe the low-energy spin dynamics of this system by an effective strongly anisotropic pseudospin-$\frac12$ model. Our ESR data show up to seven branches
of magnetic resonance in four magnetic phases arising due to subtle interplay of frustration, low dimensionality and strong anisotropy. In particular, in the low-field collinear
stripe phase, the field evolution of modes lying below 200~GHz is described reasonably good by spectra of spin-1 and spin-0 quasiparticles which we obtain using the
bond-operator technique. These well-defined excitations can be treated as conventional magnons and bound states of two magnons, respectively. In contrast, numerous excitations
lying above 200~GHz are not captured by our theory due to pronounced one-dimensional correlations inside spin chains which govern the spin dynamics at high enough energies. As
it was shown before, these modes can be most naturally interpreted as bound states of domain walls in individual chains and their sequence resembles the so-called "Zeeman
ladder" in anisotropic Izing-like spin chains. Thus, $\rm Cs_2CoBr_4$ is a system showing spin-dynamics in ordered state characteristic of both two-dimensional and
one-dimensional magnets.
\end{abstract}

\date{\today}


 \maketitle

\section{Introduction}
\label{intro}

In two-dimensional (2D) magnets, interplay of quantum fluctuations, frustration, and anisotropy leads to a variety of fascinating phenomena some of which are more characteristic
of one-dimensional (1D) systems \cite{balents,starykh}. Perhaps the most famous recent example of this kind is the exactly solvable Kitaev model in which three sorts of
bond-dependent Izing spin couplings result in spin-liquid phases with fractional spin excitations \cite{kitaev}.

\begin{figure}[t!]
\begin{center}
\vspace{0.1cm}
\includegraphics[width=0.42\textwidth]{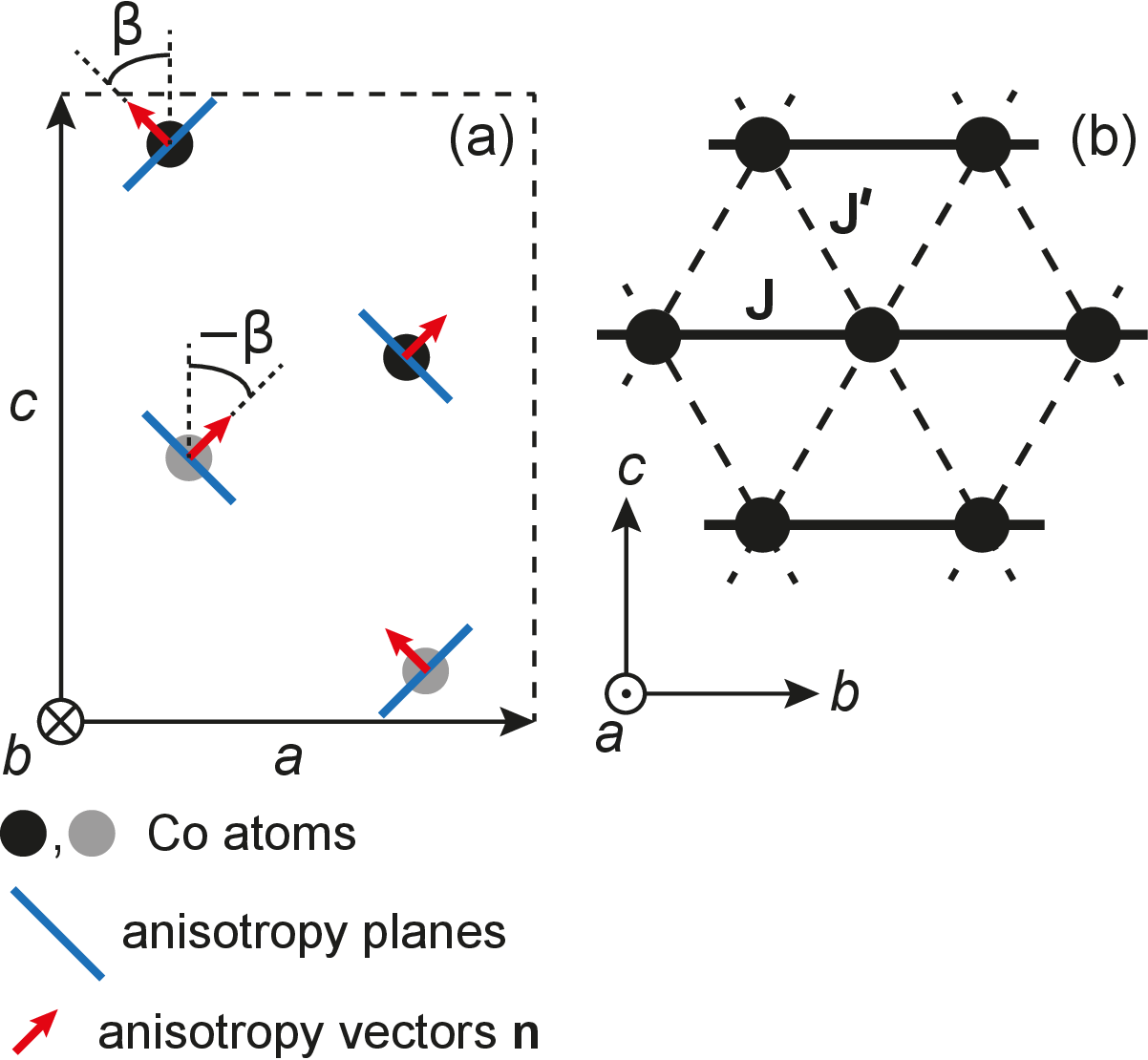}
\caption{ (a) Simplified schematic representation of the Cs$_{2}$CoBr$_{4}$ structure projected along the chain direction $b$. Dashed lines along $a$ and $c$ axes highlight the
unit cell. Black and gray dots indicate Co atoms with crystallographic positions $y= \frac{1}{4}b$ and $y= \frac{3}{4}b$, respectively. Anisotropy axes and easy planes of
Co$^{2+}$ ions are shown; $\beta\approx\pi/4$.  (b) Schematic picture of exchange paths in the $bc$ plane of Cs$_{2}$CoBr$_{4}$. \label{fig1} }
\end{center}
\end{figure}

Frustrated anisotropic quasi-2D systems can show characteristic features of 1D magnets even in phases with long-range magnetic order \cite{zvyag,kohno,starykh}. The prominent
example is Cs$_2$CuCl$_4$ demonstrating peculiar coexistence of two-spinon continuum of spin-$\frac12$ antiferromagnetic chain and quasi-2D magnons \cite{Coldea,Smirnov2,kohno}.
The latter live near the lower edge of the spinon continuum and they are captured quantitatively only in an appropriate 1D theory (in which they are treated as bound states of
two spinons) \cite{kohno,Veillette}. Cs$_2$CuCl$_4$ is described by layered spin-$\frac12$ Heisenberg model with small Dzyaloshinskii-Moriya and inter-plane interactions and
with spatially anisotropic exchange couplings within triangular planes shown in Fig.~\ref{fig1}(b). Due to the difference between $J$ and $J'\approx0.3J$, Cs$_2$CuCl$_4$ can be
viewed as an array of spin-$\frac12$ Heisenberg chains which are effectively decoupled owing to frustrating character of $J'$. Noteworthy, the 1D features survive in the
dynamics of this model up to quite large $J'\approx0.6J$ \cite{Heidarian}.

In the present paper, we address \CsCoBr which is isostructural to Cs$_2$CuCl$_4$ but which is much more anisotropic due to large spin-orbit coupling in spin-$\frac32$ Co$^{2+}$
ions.
We show that \CsCoBr also combines both quasi-2D and quasi-1D motives in its dynamics which, in contrast to Cs$_2$CuCl$_4$, are characteristic of an anisotropic system.

The orthorhombic structure of \CsCoBr corresponds to the space group P$_{nma}$ with lattice parameters $a = 10.19 \ \mathring{\textnormal{A}}$, $b = 7.73 \
\mathring{\textnormal{A}}$, $c = 13.51 \ \mathring{\textnormal{A}}$ obtained at room temperature \cite{Povarov1}. The main building blocks in the crystal are CoBr$_{4}$
distorted tetrahedra containing spin-$\frac32$ Co$^{2+}$ ions and arranged in a layered lattice with triangular $bc$ planes (which can also be viewed as an array of chains
passing along $b$ axis as in Cs$_2$CuCl$_4$). There are four tetrahedra in a crystallographic unit cell related to each other by mirror reflections in $ab$ and $bc$ planes (two
tetrahedra in each layer within a unit cell). The spin-orbit interaction in Co$^{2+}$ ions results in a strong easy-plane single-ion anisotropy with the energy parameter
$D\approx12$~K which exceeds much the intra- and inter-chain exchange couplings $J$ and $J'\approx0.45J$ (see Fig.~\ref{fig1}(b)) \cite{Povarov1,Povarov2,Povarov3}. The easy
planes are almost perpendicular to each other in two couples of Co$^{2+}$ ions in a unit cell and they intersect along a line parallel to $b$ (see Fig.~\ref{fig1}(a)). These two
anisotropies on neighboring chains effectively merge into a resultant anisotropy with easy $b$ axis so that \CsCoBr largely behaves as an axial magnet \cite{Povarov1}. The
orthorhombic symmetry does not exclude also the second axis of the anisotropy.  Due to the large single-ion anisotropy $D$, one-ion doublets $|\pm\frac12\rangle$ and
$|\pm\frac32\rangle$ are separated by the energy of $2D$ so that the consideration of the low-energy dynamics can be performed in terms of a pseudospin-$\frac12$ model
\cite{Povarov1,Povarov2,Povarov3}.

Previous  works on \CsCoBr \cite{Povarov1,Povarov2,Povarov3} revealed a long-range 3D stripe antiferromagnetic order at zero field below the N\'{e}el temperature $T_N=1.3$~K
with two collinear magnetic sublattices parallel to $b$ axis. There are also several phase transitions upon increasing of the field $\bf H$ directed along $b$ axis: the stripe
phase changes for the spin-density wave (SDW) state, then a collinear three sublattice up-up-down (UUD) phase occurs with 1/3-magnetization plateau, then a paramagnet phase with
unknown spin structure was detected, and finally the fully saturated phase arises. Neutron inelastic scattering experiments \cite{Povarov2} indicated at zero field a magnon mode
and a dispersive continuum above it similar to that observed in Cs$_2$CuCl$_4$ \cite{Coldea}. However more precise neutron studies and terahertz spectroscopy \cite{Povarov3}
show that instead of this continuum there is a sequence of at least nine dispersive excitations reminiscent of Zeeman ladder (ZL) in anisotropic weakly interacting Ising-like
chains \cite{shiba}. These modes were interpreted as bound states of domain walls (kinks or spinons) in individual chains confined by inter-plane coupling \cite{Povarov3}. It is
well known that negative roots of the Airy function describe energies of excitations at Brillouin zone (BZ) center in ZLs \cite{mccoy,Rutkevich}. It was confirmed experimentally
in $\rm CoNb_2O_6$ \cite{Coldea2}, $\rm BaCo_2V_2O_8$ \cite{Grenier,kimura}, and $\rm SrCo_2V_2O_8$ \cite{sr2,Bera}. All modes energies observed in \CsCoBr are also
quantitatively captured by roots of the Airy function except for three lower excitations lying below 0.8~meV (200~GHz) \cite{Povarov3}.

Thus, \CsCoBr is an interesting system which combines characteristic features of triangular-lattice antiferromagnets (the UUD phase with the 1/3-magnetization plateau
\cite{ChubukovGolosov}) and anisotropic quasi-1D magnets (SDW state and the ZL).

Notice that two-spinon bound states forming a ZL are linear combinations of states created from the N\'eel state by reversing several adjacent spins \cite{shiba}. They carry
spin 1 and 0 if the number of flipped spins is odd and even, respectively. Spin-1 and spin-0 excitations appear in transverse and longitudinal spin correlators, respectively.
Compared to spin-1 ZLs, spectral weight of spin-0 ZLs is generally suppressed. Then, spin-0 ZLs become apparent in neutron cross section only in systems which are not very close
to the Izing limit \cite{Grenier}. For instance, both spin-1 and spin-0 ZLs shifted relative to each other by energy were observed by unpolarized neutrons in $\rm BaCo_2V_2O_8$
\cite{Grenier}. All states in a spin-1 ZL are doubly degenerate by spin projection $S_z=\pm1$ so that they are split by magnetic field \cite{kimura,sr2}. In contrast, weak $\bf
H$ does not approximately change spin-0 ZLs. As far as we know, the magnetic field behavior of modes observed in the stripe phase of \CsCoBr was not reported before.

In the present paper, we study magnetic-field behavior of \CsCoBr in all ordered phases by the ESR measurements up to 250~GHz. Our data are complementary to previous neutron
scattering and terahertz spectroscopy results of Refs.~\cite{Povarov1,Povarov2,Povarov3} and they reveal a much richer zoo of excitations. In the stripe phase, we find numerous
weak excitations which cannot be parts of a ZL according to their field evolution (all modes are changed considerably by $\bf H$ but none of them is split by $\bf H$ as a level
in a spin-1 ZL). These excitations are described in general by our theory based on the bond operator technique (BOT) \cite{ibot}. This method is similar in spirit to the
conventional spin-wave theory but it takes into account short-range spin correlations more accurately. Besides, BOT provides an easy way to study "complex" excitations which
could appear in conventional approaches as bound states of several number of magnons. We show by the BOT that the brightest modes in the ESR spectra in the stripe phase
correspond to two spin-1 modes which behave in the field as magnons in a quasi-2D antiferromagnet with two axes of anisotropy. We identify also weaker ESR modes as well-defined
spin-0 excitations which are bound states of two magnons. The main features of our ESR data and previous inelastic neutron scattering data \cite{Povarov3} are reproduced
quantitatively below 150~GHz using the BOT. In particular, we describe quite accurately dispersion of two lower anomalies in neutron data of Ref.~\cite{Povarov3} which
correspond to the lower magnon and to the lower two-magnon bound state (which are not described by roots of the Airy function at the BZ center \cite{Povarov3}). Above 0.6~meV
($\sim150$~GHz), our theoretical results deviate from experimental data. We argue that this is due to pronounced 1D correlations in the system at large energies which cannot be
captured by the BOT. Bearing in mind also the successful attempt in Ref.~\cite{Povarov3} to identify the series of anomalies at energy greater than 0.8~meV as a ZL, we state
that \CsCoBr is a system whose spin dynamics combines characteristic features of both quasi-2D (at low enough energy) and quasi-1D (at higher energy) strongly anisotropic
magnets.

The rest of the present paper is organized as follows. We provide details of experimental setup in Sec.~\ref{exper}. Our experimental findings are discussed in
Sec.~\ref{ExpResults}. We consider the model and our theoretical approach in Sec.~\ref{theorysec}. Neutron scattering cross section observed at zero field in
Ref.~\cite{Povarov3} is described within our theory in Sec.~\ref{neutrsec}. Sec.~\ref{esrdesc} contains our theoretical interpretation of ESR spectra observed in the stripe, the
UUD, and the saturated phases. A detailed overview of all our results and conclusion can be found in Sec.~\ref{conc}.

\begin{figure}[t!]
\begin{center}
\vspace{0.1cm}
\includegraphics[width=0.35\textwidth]{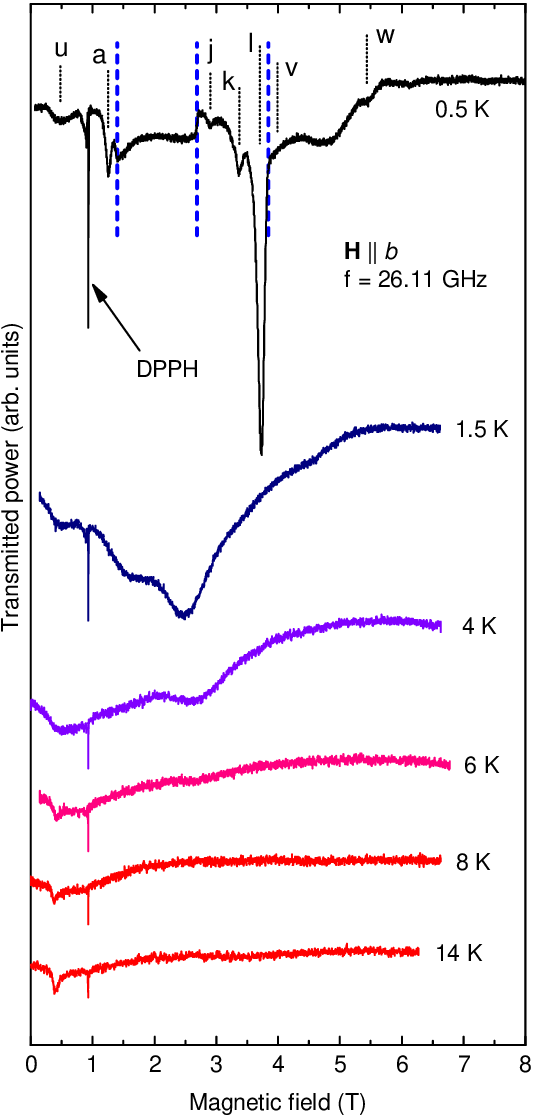}
\caption{\label{Tdep26GHz} Temperature evolution of 26.11 GHz ESR lines in \CsCoBr at ${\bf H} \parallel b$. Letters indicate modes whose frequencies are displayed on the
frequency-field diagram in Fig.~\ref{FvsHb0p5K}. Vertical dashed blue lines denote phase boundaries (see the text).}
\end{center}
\end{figure}

\begin{figure}[t!]
\begin{center}
\vspace{0.1cm}
\includegraphics[width=0.42\textwidth]{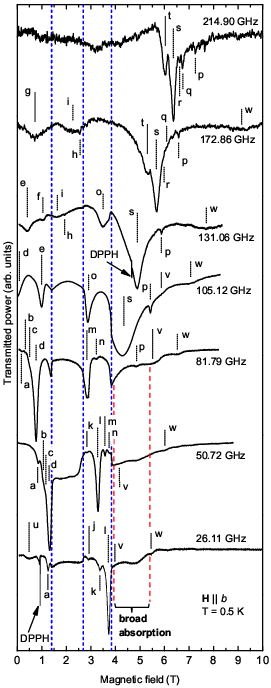}
\caption{\label{ESRlines_Hb_0p5K} ESR lines of \CsCoBr at ${\bf H} \parallel b$ at various frequencies and at $T=0.5$~K. Letters indicate modes whose frequencies are displayed
in the frequency-field diagram in Fig.~\ref{FvsHb0p5K}. Vertical dashed blue lines denote phase boundaries (see the text).}
\end{center}
\end{figure}

\clearpage

\begin{figure}[t!]
\begin{center}
\vspace{0.1cm}
\includegraphics[width=0.42\textwidth]{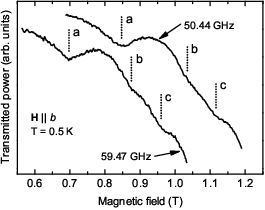}
\caption{\label{ESRlines_Hb_0p5Kmodebandc} Fragments of two ESR records in \CsCoBr at ${\bf H} \parallel b$ and $T=0.5$~K demonstrating weak modes $a$, $b$, and $c$. Letters
indicate modes whose frequencies are displayed on the frequency-field diagram in Fig.~\ref{FvsHb0p5K}.}
\end{center}
\end{figure}

\section{Experiment}
\label{exper}

The magnetic resonance signals were recorded at the fixed frequency as field dependencies of the transmitted microwave power in the frequency range 25--250~GHz using a multimode
microwave cylindrical resonators of homemade transmission-type ESR spectrometers. The latter are equipped with cryomagnets and $^3$He pump cryostat providing temperatures down
to 0.45~K and autonomous dilution microrefrigerator providing a temperature of 0.1~K \cite{Edelman}. The sample was fixed by the Apiezon grease inside the copper resonator and
placed at the maximum of the microwave magnetic field of the  TE$_{012}$ mode with the frequency 35 GHz. A small amount of 2,2-diphenyl-1-picrylhydrazyl (known as DPPH) was used
as a $g=2.00$ marker  and placed near the crystal sample. The Cs$_{2}$CoBr$_{4}$ samples studied here were from the same  batch as  in Refs.~\cite{Povarov1,Povarov2,Povarov3}.
We used samples  of different size with the mass varied from 3 to 15~mg. The sample size was chosen to be small enough  to avoid parasitic size-effects resulting in
electrodynamic resonances of samples at a given frequency.  The accuracy of the orientation of a crystal with respect to the magnetic field was checked by the X-ray
diffractometer with the accuracy of about $2^\circ$.

\section{Experimental results}
\label{ExpResults}

\begin{figure*}[t!]
\begin{center}
\vspace{0.1cm}
\includegraphics[width=0.8\textwidth]{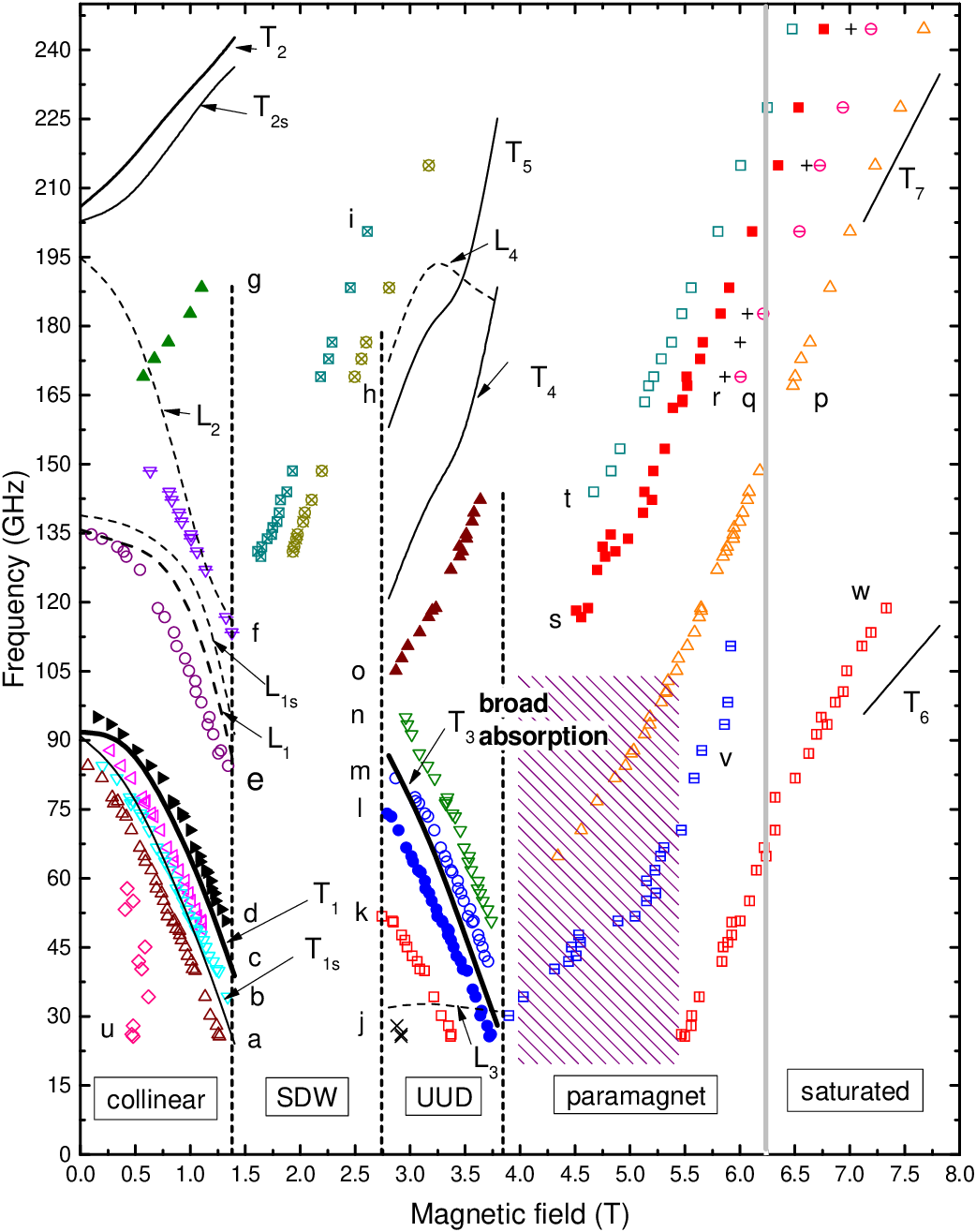}
\caption{\label{FvsHb0p5K} ESR frequency-field diagram at ${\bf H} \parallel b$  and $T=0.5$~K. Letters denote those modes
 which are indicated in Figs.~\ref{Tdep26GHz}, \ref{ESRlines_Hb_0p5K}, and \ref{ESRlines_Hb_0p5Kmodebandc}. Vertical dashed lines mark fields of
 phase transitions determined by steps on transmission records in Fig.~\ref{ESRlines_Hb_0p5K}. Grey vertical line denotes the field of the crossover to a pseudospin-saturated
phase according to Ref.\cite{Povarov1}. Solid and dashed curves
 show theoretically found spectra of spin-1 and spin-0 quasiparticles, respectively. Curves widths are proportional to spectral weights of corresponding poles in the transverse
 and in the longitudinal dynamical structure factors ${\rm Im}\chi_\perp({\bf 0},\omega)$ and ${\rm Im} \chi_\|({\bf 0},\omega)$ (see Eqs.~\eqref{chizz} and \eqref{chiperp}).
 }
\end{center}
\end{figure*}

In the low temperature range $T<1$~K our observations reveal in \CsCoBr a multi-mode ESR spectrum of ordered ground states at ${\bf H} \parallel b$. In the low-field collinear
antiferromagnetic phase we see up to 7 resonance modes at a constant field. Fig.~\ref{Tdep26GHz} illustrates the general view of spin-resonances observed at low temperature and
their temperature evolution. Records of 26.11 GHz ESR lines are presented in Fig.~\ref{Tdep26GHz} taken at ${\bf H} \parallel b $ and various temperatures. At low temperature $T=0.5$~K one can see several
intensive ESR absorption lines as well as many lines of much lower intensity. Besides, a wide range of absorption is also observable between fields 4.2~T and 5.2~T. Above the
N\'{e}el temperature, a system of wide resonances appears instead of these sharp resonances which will be described in detail elsewhere. Exploring a dense set of frequencies
we follow the field dependencies of all these modes and find the limiting fields of their existence (see examples of ESR records on different frequencies in
Fig.~\ref{ESRlines_Hb_0p5K}). In Fig.~\ref{ESRlines_Hb_0p5K}, the intensive modes are indicated by letters $l$ (already shown in Fig.~\ref{Tdep26GHz}), $d$, $g$, $o$, and $s$.
Modes $d$, $g$ belong to the collinear antiferromagnetic phase, modes $l$, $o$ exist in the UUD phase, mode $s$ is observed in the paramagnet and saturated phases. ESR lines of
much lower intensity are indicated in Figs.~\ref{Tdep26GHz} and \ref{ESRlines_Hb_0p5K} as $a$, $b$, $c$, $j$, $k$, $m$, $n$, $u$, $v$, and $w$. Taking the records on different
frequencies, we  follow the frequency-field dependencies of all these resonance lines. An example of observation of weak modes $a$, $b$, and $c$ is shown in
Fig.~\ref{ESRlines_Hb_0p5Kmodebandc}. The total number of frequencies for which the records were done is about 60 in the range 26--245 GHz. The frequency-field dependencies are
shown in Fig.~\ref{FvsHb0p5K}. The data corresponding to more intensive lines are presented by closed symbols while weak resonances are denoted by open symbols or crosses.

\begin{figure}[t!]
\begin{center}
\vspace{0.1cm}
\includegraphics[width=0.42\textwidth]{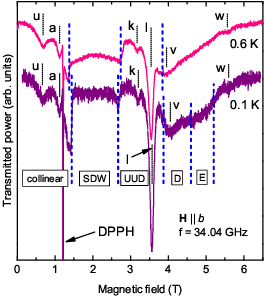}
\caption{\label{Comparison0p6and0p1K} ESR lines of \CsCoBr at 34.04~GHz, at ${\bf H} \parallel b$ and at two temperatures $T=0.6$~K and $0.1$~K. Boundaries of phases D and E are
taken from Ref.~\cite{Povarov1} (see the text).}
\end{center}
\end{figure}

\begin{figure}[t!]
\begin{center}
\vspace{0.1cm}
\includegraphics[width=0.42\textwidth]{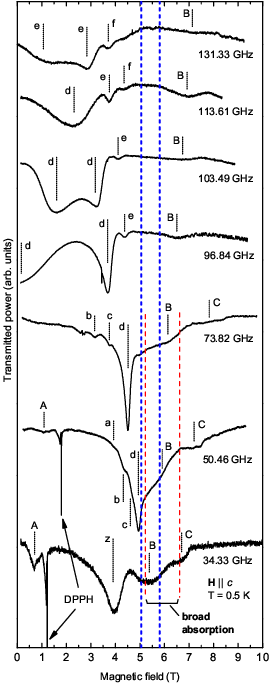}
\caption{\label{ESRlines_Hc_0p5K} ESR lines of \CsCoBr at ${\bf H} \parallel c$, at various frequencies, and at $T=0.5$~K. Letters indicate modes  whose frequencies are
displayed on the  frequency-field diagram in Fig.~\ref{FvsHc0p5K}. Thick dashed vertical lines show boundaries of phase F'' according to Ref. \cite{Povarov1}.   }
\end{center}
\end{figure}

\begin{figure}[t!]
\begin{center}
\vspace{0.1cm}
\includegraphics[width=0.42\textwidth]{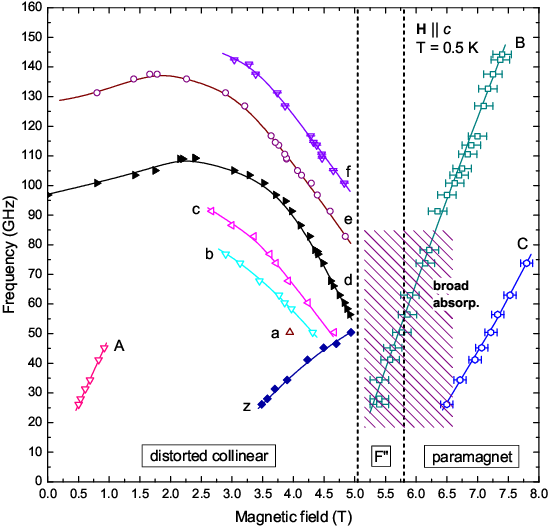}
\caption{\label{FvsHc0p5K} Frequency-field diagram of \CsCoBr at ${\bf H} \parallel c$ and at $T=0.5$~K. Intensive and weak lines are marked by closed and open symbols,
respectively. Vertical dashed lines are phase boundaries between the distorted collinear state and a phase with unknown properties denoted as F" in Ref.\cite{Povarov1}. }
\end{center}
\end{figure}

The intensive and weak modes were distinguished by a rough criterion: integral intensities of "intensive" and "weak" modes differ by more then three times. The resonance fields
are deduced as the field of local maximum absorption for intensive modes. For weak modes, the resonance fields are deduced as a field of the maximum deviation from the smooth
interpolation of the record.  Because the resonance lines are quite narrow, the error in resonance field value does not exceed the size of symbols in Fig.~\ref{FvsHb0p5K}. The
most intensive ESR line $d$ has zero-field frequency of 95 $\pm$1 ~GHz and demonstrates a tendency to fall to zero frequency  with the field increasing, which, however is broken
by a phase transition to spin-density wave phase at $H=1.4 $~T. This line has a weak lower frequency satellites $a$, $b$, $c$ with a similar falling frequency-field
dependencies. The field ranges of observed resonance modes correlate well with the phase boundaries, determined using specific heat and torque measurements in
Ref.~\cite{Povarov1}. Indeed, we observe breaks and discontinuities of frequency-field dependencies of modes at the phase boundaries at $H=1.38 \pm 0.02$~T between the stripe
collinear  and the spin-density wave phases, at $H=2.74 \pm 0.03$~T between the spin-density wave and the UUD phases, and at $H=3.6 \pm 0.02$~T between the UUD-phase and the
paramagnet phases observed in Ref.~\cite{Povarov1} (see Fig.~\ref{FvsHb0p5K}).

We have been looking also for the low temperature phase which is reported in Ref.~\cite{Povarov2} below 0.25~K in magnetic field between 4 and 6.25~T. We have probed it with
34.04~GHz microwaves at $T=0.1$~K. The record of this signal together with the record taken at $T=0.6$~K is presented in Fig.~\ref{Comparison0p6and0p1K}. We see that the low
temperature curve demonstrates a wide band of additional absorption between fields of 4 and 5.5~T, which correspond to phases "D" and "E" found in Ref. \cite{Povarov1}.

Similar to the $H$--$T$ phase diagram, the ESR spectrum of \CsCoBr in ordered states appears to be strongly anisotropic. ESR lines and frequency-field diagram
 for ${\bf H} \parallel c$ are shown in Figs.~\ref{ESRlines_Hc_0p5K} and \ref{FvsHc0p5K}, respectively.  We see that the most intensive branch has the same zero-field frequency
 95~GHz as for ${\bf H} \parallel b$. A
variety of resonance modes is observed in a field range below 5 T, while near the saturation (i.e., in the phase F") there is only resonance mode $B$ to which mode $C$ is added
in the saturated phase. Kinks in the transmission records at 73.82~GHz and 50.46~GHz indicate a phase transition at $5.05 \pm 0.03$~T from distorted collinear phase to state
$F^{\prime \prime}$ (notations of phase is from Ref. \cite{Povarov1}).

\section{Model and theoretical technique}
\label{theorysec}

\subsection{Model}

It is believed that $\rm Cs_2CoBr_4$ is described by the following spin-3/2 Hamiltonian \cite{Povarov1,Povarov2}:
\begin{eqnarray}
\label{ham0} {\cal H} &=& \sum_{i,j} \left( D \left[ \left( S_{2i,j}^x \right)^2 + \left( S_{2i+1,j}^y \right)^2 \right] + J \left( {\bf S}_{i,j}{\bf S}_{i,j+1} \right)
\right.\\
&&{} + \left. A_1 S^z_{i,j} S^z_{i,j+1} + J' \left( {\bf S}_{i,j}{\bf S}_{i+1,j} + {\bf S}_{i,j}{\bf S}_{i+1,j-1} \right)
\right.\nonumber\\
&&{} - \left. A_2 \left( S^z_{i,j} S^z_{i+1,j} + S^z_{i,j} S^z_{i+1,j-1} \right)  \right) - H \sum_{i,j} S^z_{i,j},\nonumber
\end{eqnarray}
where ${\bf S}_{i,j}$ is the $j$-th spin in the $i$-th chain passing along $b$ axis (see Fig.~\ref{fig1}(b)), $J>0$ and $J'>0$ are intra- and inter-chain exchange coupling
constants, respectively, $A_1\ll J$ and $A_2\ll J'$ are small anisotropies, $D\gg J,J'$ is the easy-plane anisotropy, a small interaction between triangular planes is neglected,
$z$ and $b$ axes are parallel to each other, $x$ and $y$ axes are mutually orthogonal and they are parallel to hard axes in neighboring chains shown in Fig.~\ref{fig1}(a).
Notice that the easy plane alternates from chain to chain between $xz$ and $yz$ planes. As compared with the model suggested in Refs.~\cite{Povarov1,Povarov2}, we introduce only
the small anisotropy $A_2$ in Eq.~\eqref{ham0}. It was proposed in Refs.~\cite{Povarov1,Povarov2} based on the linear spin-wave theory and a mean-field analysis that
$D\approx1.2$~meV, $J\approx0.2$~meV, $J'\approx0.4 J$, and $A_1\approx0.1J$.

In the local coordinate frame in which the quantized axis is directed perpendicular to the easy plane, the lower doublet |$\pm$1/2$\rangle$ of each spin is separated
considerably from the upper doublet |$\pm$3/2$\rangle$ due to large $D$. This allows to discuss the low-energy dynamics on a simpler model by introducing pseudospin-1/2 ${\bf
s}_{i,j}$ at each lattice site which describes the lower doublet |$\pm$1/2$\rangle$. The transition from spins to pseudospins can be made according to the rule which readily
follows from comparison of matrix elements of spin operators 1/2 and 3/2:
\begin{equation}
\label{rule}
\begin{aligned}
S^x_{2i,j}&\mapsto s^x_{2i,j}, & S^y_{2i,j}&\mapsto 2s^y_{2i,j}, &
S^z_{2i,j}&\mapsto 2s^z_{2i,j}, \\
S^x_{2i+1,j}&\mapsto 2s^x_{2i+1,j}, & S^y_{2i+1,j}&\mapsto s^y_{2i+1,j}, & S^z_{2i+1,j}&\mapsto 2s^z_{2i+1,j}.
\end{aligned}
\end{equation}
As a result, one comes from Eq.~\eqref{ham0} to the following pseudospin-1/2 Hamiltonian:
\begin{eqnarray}
\label{hamps} {\cal H}_{ps} &=& \sum_{i,j} \left( 4J \left( {\bf s}_{2i,j}{\bf s}_{2i,j+1} \right) - 3J s^x_{2i,j} s^x_{2i,j+1}
\right.\nonumber\\
&&{}
 + \left. 4J \left( {\bf s}_{2i+1,j}{\bf s}_{2i+1,j+1} \right) - 3J s^y_{2i+1,j} s^y_{2i+1,j+1}
\right.\nonumber\\
&&{} + \left. 4A_1 s^z_{i,j} s^z_{i,j+1} + 2J' \left( {\bf s}_{i,j}{\bf s}_{i+1,j} \right)
\right.\nonumber\\
&&{} + \left. 2J' \left( {\bf s}_{i,j}{\bf s}_{i+1,j-1} \right) + \left( 2J' - 4A_2 \right) s^z_{i,j} s^z_{i+1,j}
\right.\nonumber\\
&&{} \left. + \left( 2J' - 4A_2 \right) s^z_{i,j} s^z_{i+1,j-1} \right) - 2H \sum_{i,j} s^z_{i,j}.
\end{eqnarray}

At zero field, model \eqref{hamps} can be simplified further by performing the canonical transformation
\begin{equation}
\label{rot}
    s^x_{2i,j} = -\tilde s^{ y}_{2i,j},
    \quad
    s^y_{2i,j} = -\tilde s^{ x}_{2i,j},
    \quad
    s^z_{2i,j} = -\tilde s^{ z}_{2i,j}
\end{equation}
which is a result of subsequent rotations by $\pi/2$ around $z$ axis and by $\pi$ around $y$ axis in chains with even numbers. After this transformation, all chains become
equivalent and the Hamiltonian acquires the form
\begin{eqnarray}
\label{hamh0} {\cal H}_{ps}^{H=0} &=& \sum_{i,j} \left( 4J \left( \tilde {\bf s}_{i,j}\tilde {\bf s}_{i,j+1} \right) - 3J \tilde s^y_{i,j} \tilde s^y_{i,j+1}
\right.\nonumber\\
&&{} + \left. 4A_1 \tilde s^z_{i,j} \tilde s^z_{i,j+1}
 - 2J' \left( \tilde s^x_{i,j}\tilde s^y_{i+1,j} + \tilde s^y_{i,j}\tilde s^x_{i+1,j} \right)
\right.\nonumber\\
&&{}- \left. 2J' \left( \tilde s^x_{i,j}\tilde s^y_{i+1,j-1} + \tilde s^y_{i,j}\tilde s^x_{i+1,j-1} \right)
\right.\nonumber\\
&&{}
 - \left. 4\left(J' - A_2\right)
\left( \tilde s^z_{i,j} \tilde s^z_{i+1,j} + \tilde s^z_{i,j} \tilde s^z_{i+1,j-1} \right)
 \right).
\end{eqnarray}
Notice that there is no such canonical transformation making chains equivalent at finite field.

As it is demonstrated below and as it was observed before experimentally \cite{Povarov1,Povarov2,Povarov3}, due to anisotropic spin interactions, model \eqref{hamps} shows the
stripe and the UUD collinear phases at small fields with sublattices magnetizations directed along $z$ axis and with gapped spectra. If Hamiltonian \eqref{hamps} was invariant
with respect to rotation around $z$ axis, all system states would be characterized by projection of the total spin $S_z$. In particular, there would be magnetization plateaus in
the stripe and in the UUD phases, all excitations could be characterized by their $S_z$ values (e.g., spin-0, spin-1 excitations), and poles of transverse and longitudinal (with
respect to $z$ axis) dynamical spin susceptibilities would be determined only by spin-1 and spin-0 quasiparticles, respectively. But, formally, Hamiltonian \eqref{hamps} is not
invariant with respect to rotation around $z$ axis due to $s^xs^x$ and $s^ys^y$ terms (i.e., due to two types of chains). However, it was observed experimentally and
theoretically (see below) that magnetization plateaus in the stripe and in the UUD phases do exist and our calculations below do demonstrate the decoupling of the longitudinal
and the transverse channels. This interesting situation arises in this model due to an effective merging of $s^xs^x$ and $s^ys^y$ terms on neighboring chains into a single
"mean" anisotropy with $z$ easy axis.

We calculate below dynamical spin susceptibilities
\begin{eqnarray}
\label{chi} \chi_{\alpha\beta}({\bf k},\omega) &=& i\int_0^\infty dt e^{i\omega t}
\left\langle \left[ S^\alpha_{\bf k}(t), S^\beta_{-\bf k}(0) \right] \right\rangle,\\
\label{chizz}
\chi_\|({\bf k},\omega) &=& \chi_{zz}({\bf k},\omega),\\
\label{chiperp} \chi_\perp({\bf k},\omega) &=& \chi_{xx}({\bf k},\omega) +  \chi_{yy}({\bf k},\omega)
\end{eqnarray}
in models \eqref{hamps} and \eqref{hamh0} using relation \eqref{rule} and describe our ESR and previous inelastic neutron scattering data \cite{Povarov3}. Poles of
$\chi_{\alpha\beta}({\bf k},\omega)$ determine spectra of spin excitations in the system. We use below the bond-operator technique (BOT) proposed in Ref.~\cite{ibot}.

\subsection{Bond operator technique for spin-$\frac12$ models}
\label{bot}

\begin{figure}
\includegraphics[scale=0.4]{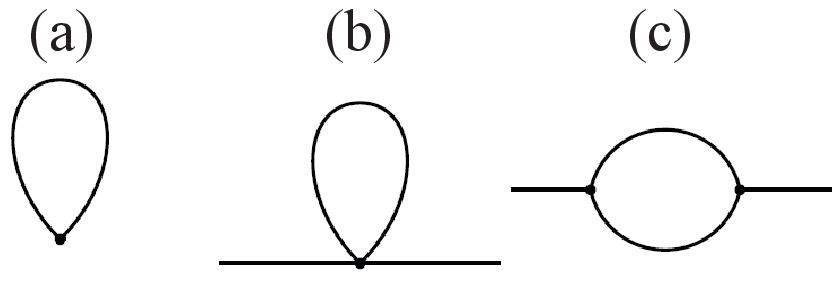}
\caption{Diagrams giving corrections of the first-order in $1/n$ to (a) the ground state energy and the staggered magnetization, and (b), (c) to self-energy parts. \label{diag}}
\end{figure}

The main idea of this approach is to take into account all spin degrees of freedom in the magnetic unit cell containing several spins 1/2 by building a bosonic spin
representation reproducing the spin commutation algebra. A general scheme of construction of such representation for arbitrary number of spins in the unit cell is described in
detail in Ref.~\cite{ibot}. We consider now briefly the main steps of this procedure by the example of four spins in the unit cell which is relevant for the stripe phase in
model \eqref{hamps}. First, we introduce 15 Bose operators in each unit cell which act on 16 basis functions of four spins $|0\rangle$ and $|e_i\rangle$ ($i=1,...,15$) according
to the rule
\begin{equation}
\label{bosons}
    a_i^\dagger |0\rangle = |e_i\rangle, \quad i=1,...,15,
\end{equation}
where $|0\rangle$ is a selected state playing the role of the vacuum. Then, we build the bosonic spin transformation in the unit cell as it is described in Ref.~\cite{ibot}
which turns out to be quite bulky and which is presented in Ref.~\cite{ibot}. The code in the Mathematica software which generates this representation is also available in
Ref.~\cite{supp2}. There is a formal artificial parameter $n$ in this spin transformation that appears in operator $\sqrt{n-\sum_{i=1}^{15}a_i^\dagger a_i}$ by which linear in
Bose operators terms are multiplied (cf.\ the term $\sqrt{2S-a^\dagger a}$ in the Holstein-Primakoff representation). It prevents mixing of states containing more than $n$
bosons and states with no more than $n$ bosons (then, the physical results of the BOT correspond to $n=1$). Besides, all constant terms in our spin transformation are
proportional to $n$ whereas bilinear in Bose operators terms do not depend on $n$ and have the form $a_i^\dagger a_j$. We introduce also separate representations via operators
\eqref{bosons} for terms ${\bf s}_i{\bf s}_j$, $s_i^xs_j^x$, $s_i^ys_j^y$, and $s_i^zs_j^z$ in the Hamiltonian in which $i$ and $j$ belong to the same unit cell. Constant terms
in these representations are proportional to $n^2$ and terms of the form $a_i^\dagger a_j$ are proportional to $n$ \cite{ibot}. Thus, we obtain a close analog of the
conventional Holstein-Primakoff spin transformation which reproduces the commutation algebra of all spin operators in the unit cell for all $n>0$ and in which $n$ is the
counterpart of the spin value $S$. In analogy with the spin-wave theory (SWT), expressions for observables are found in the BOT using the conventional diagrammatic technique as
series in $1/n$. This is because terms in the Bose-analog of the spin Hamiltonian containing products of $i$ Bose operators are proportional to $n^{2-i/2}$ (in the SWT, such
terms are proportional to $S^{2-i/2}$). For instance, to find the ground-state energy, the staggered magnetization, and self-energy parts in the first order in $1/n$ one has to
calculate diagrams shown in Fig.~\ref{diag} (as in the SWT in the first order in $1/S$).

Notice that states $|0\rangle$ and $|e_i\rangle$ in Eq.~\eqref{bosons} are linear combinations of elementary quantum states which are simple products of states
$|\uparrow\rangle$ and $|\downarrow\rangle$ of each spin in the unit cell. Coefficients in these combinations contain parameters which should be found by minimizing the term in
the Hamiltonian not containing Bose operators (linear in Bose operator terms in the Hamiltonian vanish at this set of parameters) \cite{ibot}. In particular, we determine in
this way properties of the ground state (e.g., the spin ordering) in the zeroth order in $1/n$.

Previous applications of the BOT to two-dimensional spin-$\frac12$ systems well studied before both theoretically (by other numerical and analytical methods) and experimentally
show that first $1/n$ terms in most cases give the main corrections to renormalization of observables if the system is not very close to a quantum critical point
\cite{ibot,aktersky,iboth,itri,itrih,itrij1j2}. Similarly, first $1/S$ corrections in the SWT frequently make the main quantum renormalization of observable quantities even at
$S=1/2$, see, e.g., Ref.~\cite{monous}.
 Importantly, because the spin commutation algebra is reproduced in our method at any $n>0$, the proper number of Goldstone
excitations arises in ordered phases in any order in $1/n$ (unlike the vast majority of other versions of the BOT proposed so far \cite{ibot}).

As quantum spin correlations inside the unit cell are taken into account accurately within the BOT, we achieved quite precise description of salient features of short-wavelength
quasiparticles in different systems some of which cannot be described even qualitatively using conventional analytical approaches \cite{iboth,itri,itrih,itrij1j2}. Although the
BOT is technically very similar to the SWT, the main disadvantage of this technique is that it is very bulky (e.g., the part of the Hamiltonian bilinear in Bose operators
contains more than 100 terms) and it requires time-consuming numerical calculation of diagrams.

The three-spin variant of the BOT can be constructed in a similar way. It was  used in Refs.~\cite{itri,itrih,itrij1j2} for consideration of various ordered phases in
triangular-lattice antiferromagnets. The code in the Mathematica software which generates this spin transformation can be found in Ref.~\cite{supp1}. We employ below the
three-spin variant of the BOT for discussion of the UUD state and the four-spin variant will be utilized for consideration of the stripe and the saturated phases.

Notice that the BOT allows to consider numerous complex excitations which can arise in standard approaches as bound states of conventional quasiparticles (magnons or triplons)
\cite{ibot}. For instance, there are only four bosons in the four-spin BOT which describe conventional magnons in the stripe phase. The remaining 11 bosons are responsible for
some other excitations which are built on quantum states of the whole unit cell (see Eq.~\eqref{bosons}). In common methods, discussion of the bound states requires analysis of
some infinite series of diagrams for vertexes. As a result, spectra of bound states cannot be obtained as series of some parameter within conventional approaches because it is
normally impossible to take into account all the required diagrams. In contrast, the existence of separate bosons in BOTs (describing some bound states in considered models)
allows to find their spectra as series in $1/n$ by calculating the same diagrams as for the common quasiparticles (e.g., diagrams shown in Figs.~\ref{diag}(b) and \ref{diag}(c)
in the first order in $1/n$). For instance, in the ordered phase of the isotropic square-lattice antiferromagnet, the version of the BOT with two-site unit cell contains three
bosons describing  two spin-1 excitations (conventional magnons) and one spin-0 quasiparticle (the Higgs mode) \cite{ibot}. The price to pay for the increasing of the
quasiparticles zoo is the bulky theory.

In the present study, we discuss the stripe phase at zero field (model \eqref{hamh0}) using four-spin BOT with two variants of the unit cell: the plaquette-like unit cell
containing two couples of spins from neighboring chains and the chain-like unit cell with four spins from the same chain (see insets in Fig.~\ref{dsf1212}). This allows us to
reveal the role of 1D correlations in the considered model on conventional spin-1 excitations (magnons) and on spin-0 quasiparticles (which could arise in the SWT as bound
states of two magnons) appearing in the longitudinal channel.

We take into account below diagrams shown in Fig.~\ref{diag}(b) and \ref{diag}(c) to find all self-energy parts $\Sigma(\omega,{\bf k})$ in the first order in $1/n$. We use
(bare) Green's functions of the harmonic approximation in these calculations. Spectra of elementary excitations are obtained by finding zeroes of Green's functions denominators
taking into account $\omega$-dependence of self-energy parts. This scheme of calculations proved to give an accurate description of elementary excitations some of which had no
counterparts not only in the semiclassical spin-wave theory but also in the harmonic approximation of the BOT \cite{iboth,itri,itrih,itrij1j2}.

\section{Theoretical description of previous neutron data at $H=0$}
\label{neutrsec}

Let us apply the BOT described above to $\rm Cs_2CoBr_4$. We have achieved the best fit of previously obtained neutron data \cite{Povarov3} at $H=0$ and our ESR spectra (see the
next section) with the following parameters in pseudospin model \eqref{hamps}:
\begin{equation}
\label{param}
\begin{aligned}
    J &= 0.165~{\rm meV}, & A_1 &= 0.34J,\\
    J' &= 0.45J, &  A_2 &= 0.1J'.
\end{aligned}
\end{equation}
As compared with previous considerations of $\rm Cs_2CoBr_4$ using the linear SWT and a mean-field approach \cite{Povarov1,Povarov2,Povarov3}, we have increased $A_1$ by 3.4
times, slightly reduced $J$, and introduced small $A_2$.

\begin{figure}
\includegraphics[scale=1.0]{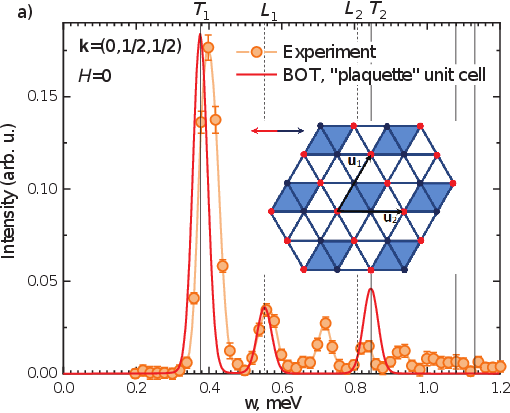}
\includegraphics[scale=1.0]{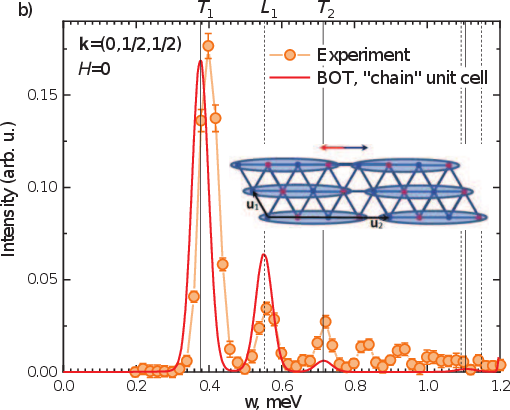}
\caption{Inelastic neutron scattering data at ${\bf k}=(0,0.5,0.5)$ in zero field taken from Fig.~1(a) of Ref.~\cite{Povarov3} and dynamical structure factor (DSF)
\eqref{neutron} calculated within the BOT in the first order in $1/n$ with (a) plaquette-like and (b) chain-like unit cells shown in insets (where lattice sites belonging to two
collinear sublattices are distinguished by color). Theoretical results are integrated in the interval $(-5,5)$ of the momentum component perpendicular to triangular planes while
experimental data are "fully integrated" along that direction \cite{Povarov3}. Theoretical results are convoluted with the experimental energy resolution of 0.02~meV. Dashed and
solid vertical lines mark positions of peaks obtained theoretically in the longitudinal and transverse DSFs, respectively (i.e., energies of well-defined spin-0 and spin-1
excitations). Spin-1 and spin-0 excitations marked as $T_{1,2}$ and $L_{1,2}$ are discussed in detail in the text. \label{dsf1212}}
\end{figure}

In zero field, we can perform canonical transformation \eqref{rot} and consider model \eqref{hamh0} in which all chains are equivalent. This allows to apply two modifications of
the four-spin BOT with two kinds of the unit cell to reveal the character of low-energy excitations observed recently in neutron experiment \cite{Povarov3} and in our ESR
measurements. To describe available neutron data, we calculate the following dynamical structure factor \cite{Lowesey}:
\begin{equation}
\label{neutron} {\cal S}({\bf k},\omega) = \frac1\pi  {\rm Im} \sum_{\alpha,\beta} g_{\alpha}g_{\beta} \left( \delta_{\alpha\beta} - \widehat k_\alpha \widehat k_\beta \right)
\chi_{\alpha\beta}({\bf k},\omega),
\end{equation}
where $\alpha,\beta=x,y,z$, $\widehat{\bf k}={\bf k}/k$, $\chi_{\alpha\beta}({\bf k},\omega)$ are given by Eq.~\eqref{chi}, and we use that $g$-tensor is diagonal and has
components $g_{x}=2.37(1)$, $g_{y}=2.42(1)$, and $g_{z}=2.47(2)$  \cite{Povarov1}. For the plaquette-like and the chain-like unit cells shown in insets of Fig.~\ref{dsf1212},
${\bf S}_{\bf k}$ in Eq.~\eqref{chi} has the form
\begin{equation}
\label{skstripe}
    {\bf S}_{\bf k} = \frac{1}{2}
    \left(
    {\bf S}_{1\bf k} + {\bf S}_{2\bf k}e^{-ik_2/2} + {\bf S}_{3\bf k}e^{-i(k_1+k_2)/2} + {\bf S}_{4\bf k}e^{-ik_1/2}
    \right),
\end{equation}
and
\begin{equation}
\label{skchain}
    {\bf S}_{\bf k} = \frac{1}{2}
    \left(
    {\bf S}_{1\bf k} + {\bf S}_{2\bf k}e^{-ik_1/4} + {\bf S}_{3\bf k}e^{-ik_1/2} + {\bf S}_{4\bf k}e^{-i3k_1/4}
    \right),
\end{equation}
where ${\bf k} = k_1{\bf f}_1 + k_2{\bf f}_2$, ${\bf f}_{1,2}$ are reciprocal lattice vectors corresponding to translation vectors ${\bf u}_{1,2}$ in the real space depicted in
insets of Fig.~\ref{dsf1212}, and ${\bf S}_i$ is the $i$-th spin operator in the unit cell which is related with pseudospin $\tilde{\bf s}_i$ via Eqs.~\eqref{rule} and
\eqref{rot}.

It is argued in Ref.~\cite{Povarov3} that chains are effectively decoupled in the mean-field-RPA approximation at momenta with $k_b=0,\pi$ due to frustration inherent to
triangular-lattice antiferromagnets. As a result, a sequence of at least 9 peaks appears in the neutron cross section at ${\bf k}=(0,1/2,1/2)$ whose intensity decrease upon the
energy increasing (see Fig.~\ref{dsf1212}). These peaks are interpreted in Ref.~\cite{Povarov3} as a signature of bound states of domain walls inside the effectively isolated
chains having strong easy-axis anisotropy (reminiscent of ZLs in quasi-1D Ising-like systems \cite{shiba}). The attraction between domain walls stabilizing the bound states is
attributed to small inter-plane interaction which is neglected in our theoretical consideration. It should be noted that three lower peaks are not described accurately enough
within this line of arguments (see Fig.~1(d) in Ref.~\cite{Povarov3}). We show now that the BOT describes well two lower peaks. Besides, the BOT shows that these peaks
correspond to spin-1 and spin-0 excitations which cannot be a part of the same ZL. Notice that both spin-1 and spin-0 quasiparticles are probed in Ref.~\cite{Povarov3} by
unpolarized neutrons whose cross section \eqref{neutron} is a mixture of longitudinal and transverse DSFs. The THz spectroscopy used in Ref.~\cite{Povarov3} also probed both the
longitudinal and the transverse channels because the unpolarized radiation propagated along the crystallographic $c$ axis in that experiment.

Results of our calculations within the BOT are also presented in Fig.~\ref{dsf1212}, where dashed and solid vertical lines mark positions of peaks obtained theoretically in the
longitudinal and transverse DSFs, respectively. Due to the large anisotropy producing large gaps in all spectra of excitations, all the observed quasiparticles are well-defined.
It is easy to ascertain within the BOT what kind of excitation of the unit cell describes each boson (at least in the harmonic approximation) \cite{ibot,itri,itrih}. Spin-1 and
spin-0 quasiparticles marked as $T_1$ and $L_1$ in both Figs.~\ref{dsf1212}(a) and \ref{dsf1212}(b) correspond to a single spin flip and to flips of two neighboring spins inside
the same chain, respectively. Notice that anomalies corresponding to these quasiparticles describe well experimental data in both versions of the BOT. Then, it can be said that
$T_1$ is a conventional magnon (which was also obtained within the SWT in Ref.~\cite{Povarov2}) and $L_1$ is a two-magnon bound state.

Elementary excitation marked as $T_2$ in Fig.~\ref{dsf1212} is also built on single spin flips but the inclusion is also significant of spin flips of three neighboring spins
inside the same chain which are taken into account more accurately in the first order in $1/n$ in the chain-like geometry. That is why the position and the height of the anomaly
corresponding to $T_2$ differ noticeably in Figs.~\ref{dsf1212}(a) and \ref{dsf1212}(b).

Spin-0 quasiparticle $L_2$ denoted in Figs.~\ref{dsf1212}(a) corresponds to flipping of two nearest spins in neighboring chains. Its spectral weight is very small (that is why
it is invisible in the neutron data) and it appears only in the version of the BOT with the plaquette-like unit cell. In the "chain" geometry, it could appear only as a bound
state of other bosons introduced in this version of the BOT (which we do not consider here). Then, the existence of this excitation which is not a free-chain-like quasiparticle
enriches the dynamical properties of the considered system. The tiny spectral weight of this spin-0 quasiparticle can be increased in \CsCoBr by a small spin interaction not
taken into account in the considered model which mixes longitudinal and the transverse channels (e.g., small Dzyaloshinsky-Moria interaction, some other anisotropy, or an
accidental small magnetic field component transverse to $b$ axis).

It is seen from Fig.~\ref{dsf1212} that BOT is unable to describe quantitatively other small anomalies seen in neutron data at $\omega>0.6$~meV. This may be an additional
indication that profound one-dimensional correlations are responsible for their properties as it is proposed in Ref.~\cite{Povarov3}.

\begin{figure*}
\includegraphics[scale=0.9]{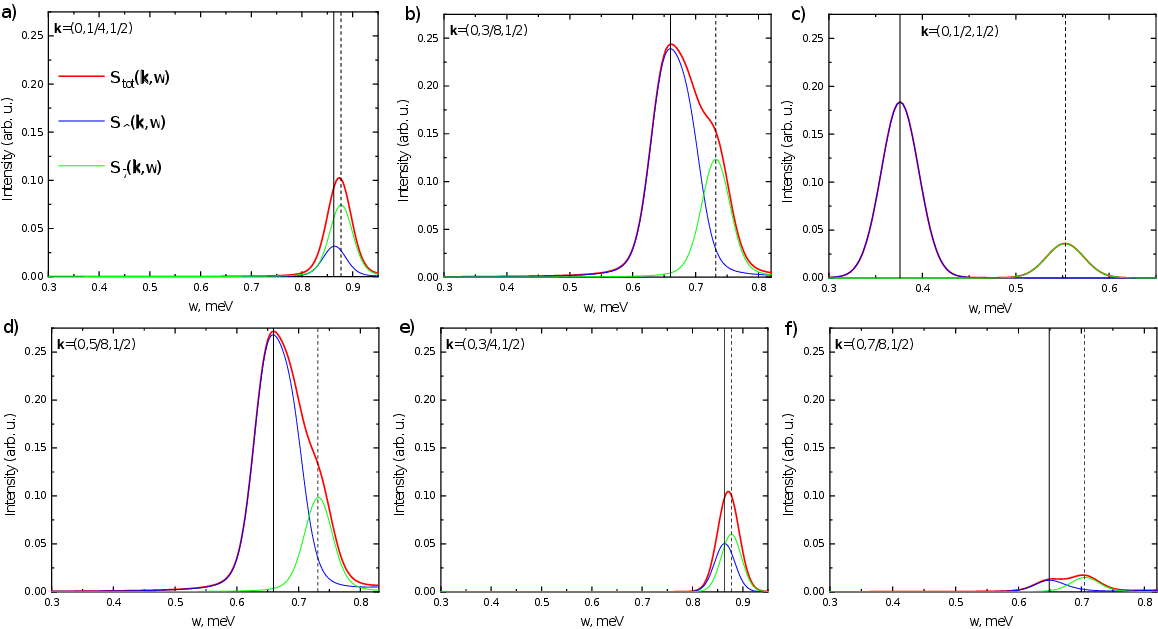}
\caption{ (a)--(f) Dynamical structure factor (DSF) ${\cal S}_{tot}({\bf k},\omega)$ given by Eq.~\eqref{neutron} and calculated within the BOT in the first order in $1/n$ with
the plaquette-like unit cell at selected momenta. Contributions to ${\cal S}_{tot}({\bf k},\omega)$ of transverse ${\cal S}_{\perp}({\bf k},\omega)$ and longitudinal ${\cal
S}_\|({\bf k},\omega)$ DSFs are also shown. The neutron cross section is integrated in the interval $(-5,5)$ of the momentum component perpendicular to triangular planes. Dashed
and solid vertical lines mark positions of peaks obtained theoretically in the longitudinal and transverse DSFs, respectively (i.e., energies of well-defined spin-0 and spin-1
excitations denoted as $L_1$ and $T_1$ in Fig.~\ref{dsf1212}). Theoretical results are convoluted with the energy resolution of 0.02~meV (inherent to the experimental data
presented in Fig. 3a of Ref. \cite{Povarov3}). The maxima of ${\cal S}_{tot}({\bf k},\omega)$, the shape of its $\omega$-dependence and the integral intensity correspond well to
the experimental data in Fig.3(a) of Ref. \cite{Povarov3}, where a false color plot is presented of the neutron scattering intensity. \label{dsfall}}
\end{figure*}

Fig.~\ref{dsfall} shows two lower anomalies (marked as $T_1$ and $L_1$ in Fig.~\ref{dsf1212}) in neutron scattering intensity ${\cal S}_{tot}({\bf k},\omega)$ given by
Eq.~\eqref{neutron} and calculated within the BOT in the first order in $1/n$ with the plaquette-like unit cell for six momenta. There is a good agreement of Fig.~\ref{dsfall}
with neutron data presented in Fig.~3(a) of Ref. \cite{Povarov3} in energies, shapes, and magnitudes of two lower maxima.

We demonstrate below that all quasiparticles just discussed (i.e., $T_{1,2}$ and $L_{1,2}$) are observed also in our ESR experiment in the stripe phase at finite field. Besides,
we obtain also low-intensive excitations looking as close satellites of $T_{1,2}$ and $L_1$ which are indistinguishable in Figs.~\ref{dsf1212} and \ref{dsfall} and which are not
mentioned in this section.

\section{Theoretical description of ESR results at ${\bf H}\|b$ and discussion}
\label{esrdesc}

We find using the BOT that the ground state of model \eqref{hamps} with parameters \eqref{param} obeys the stripe and the UUD spin orderings within experimentally found
boundaries of these phases. The longitudinal magnetization is obtained to be equal to 0 and $M_s/3$ in these two states, respectively, where $M_s$ is the saturation
magnetization \footnote{ To be precise, the longitudinal magnetization per spin increases upon the field increasing from 0 to 0.006 in the stripe phase and by 0.7\% in the UUD
state. }. It is difficult to discuss within the BOT transitions to the SDW phase from the stripe and from the UUD states because the SDW has an incommensurate ordering and
because these transitions are of the first-order type as our ESR measurements indicate. We find that the transition from the UUD state to the paramagnet phase takes place near
4~T (in agreement with experiment) as a result of a "condensation" of the lower magnon mode at an incommensurate momentum. The transition from the saturated state (in which all
pseudospins are oriented along the field) to the paramagnet one occurs upon "condensation" of spin-2 excitations. However, the spectrum of spin-2 excitations acquires great
renormalization in the first order in $1/n$ that makes our technique unsuitable for discussion of this phase transition (one has to go beyond the first order in $1/n$ which is a
difficult task). Then, we do not consider theoretically in detail phase transitions (and do not determine boundaries of phases) and focus on spectra inside the stripe, the UUD,
and the saturated phases within their experimentally determined boundaries.

\subsection{Stripe phase}

Both transverse and longitudinal polarizations of microwave field with respect to the external field are present in our ESR experiment because the sample size is comparable to
the length of the electromagnetic wave. That is why both spin-0 and spin-1 excitations can appear in our ESR measurements. Besides, spin-0 excitations from the longitudinal
channel can arise in our experiment also due to a small anisotropic interaction and/or a small transverse magnetic field component.

We plot in Fig.~\ref{FvsHb0p5K} the theoretical spectra at $\bf k=0$ of spin-1 and spin-0 quasiparticles by solid and dashed lines, respectively, which we obtain within the
four-spin BOT with the plaquette-like unit cell. Spin-1 modes are denoted as $T_1$, $T_2$, $T_{1s}$, $T_{2s}$ and spin-0 excitations are marked as $L_1$, $L_2$, $L_{1s}$. Line
widths in Fig.~\ref{FvsHb0p5K} are proportional to spectral weights of corresponding poles in the transverse and in the longitudinal dynamical structure factors ${\rm
Im}\chi_\perp({\bf 0},\omega)$ and ${\rm Im} \chi_\|({\bf 0},\omega)$ (see Eqs.~\eqref{chizz} and \eqref{chiperp}).

It is seen from Fig.~\ref{FvsHb0p5K} that lines $T_1$, $T_2$, and $L_1$ obtained within the BOT have much less intense satellites $T_{1s}$, $T_{2s}$, and $L_{1s}$. The origin of
satellites of two spin-1 excitations can be explained as follows. Notice first that $T_1$ and $T_2$ excitations correspond to magnons which appear also within the linear SWT (as
in Ref.~\cite{Povarov2} considering model \eqref{hamps}), where they are both doubly degenerate due to four spins in the magnetic unit cell. These spin-1 lines, are split by
quantum fluctuations within the BOT resulting in $T_{1s}$ and $T_{2s}$ satellites. Such modes splitting was observed by the BOT also in other models \cite{itrij1j2,itri} (the
splitting of magnon spectrum obtained in Ref.~\cite{itri} at point $M$ of the BZ in the triangular-lattice Heisenberg model is in agreement with experiment).

Notice that the crystallographic momentum (0,1/2,1/2) discussed above in the neutron experiment corresponds to (0,1,1) momentum (which, in turn, is equivalent to $\bf k=0$) in
the considered BOT with the plaquette-like unit cell. That is why modes $T_{1,2}$ and $L_{1,2}$ are the same in Figs.~\ref{FvsHb0p5K} and \ref{dsf1212}. Weak satellites $T_{1s}$
and $T_{2s}$ predicted by the BOT in the first order in $1/n$ are too close to the intense $T_{1,2}$ to be resolved in the neutron experiment at $H=0$.

It is seen from Figs.~\ref{dsf1212} and \ref{dsfall} that $T_1$ and $L_1$ excitations observed in the BOT describe quite accurately two lower neutron anomalies and modes $d$ and
$e$ in ESR data in Fig.~\ref{FvsHb0p5K}. However, the lower strong signal $d$ obtained experimentally has three less intense satellites $a$, $b$, and $c$ instead of one $T_{1s}$
predicted by the theory. We have no explanation for this discrepancy at the moment. Extra satellites may be a result of small spin interactions not included in our model. It is
also seen from Fig.~\ref{FvsHb0p5K} that weak satellite mode $L_{1s}$ is not detected experimentally.

The upper intense spin-1 mode shown in Fig.~\ref{FvsHb0p5K} corresponds to $T_2$ quasiparticle in Fig.~\ref{dsf1212}. As it is discussed above, the spectrum of this mode is
renormalized downward considerably by strong in-chain spin fluctuations which are not taken into account properly in the first order in $1/n$ in the version of the BOT with
plaquette-like unit cell. At $H=0$, the downward renormalization arises from $0.84~{\rm meV}\approx 203$~GHz to $0.72~{\rm meV}\approx 174$~GHz (see Fig.~\ref{dsf1212}). A
simple shift down of $T_2$ curve by 30~GHz brings it quite close to $g$ mode observed in the ESR experiment (see Fig.~\ref{FvsHb0p5K}). Then, we relate mode $g$ to $T_2$.

The upper spin-0 mode in Fig.~\ref{FvsHb0p5K} corresponds to $L_2$ elementary excitation in Fig.~\ref{dsf1212}(a) (i.e., to the bound state of two magnons in neighboring
chains). It is seen from Fig.~\ref{FvsHb0p5K} that its spectrum is very close below 150~GHz to weak $f$ mode observed experimentally.

It should be noted also that despite a first-glance chaotic variety of ESR modes, two bright branches $d$ and $g$ going down and up upon the field increasing are similar to the
spectrum of a collinear antiferromagnet in magnetic field directed along its easy axis. The falling branch of the conventional antiferromagnet should soften at the spin-flop
field. However, the spin-flop transition does not occur in \CsCoBr because of the earlier transition to the SDW phase. This results in finite frequency of mode $d$ at the right
boundary of the stripe phase in Fig.~\ref{FvsHb0p5K}. In the conventional antiferromagnet with two axes of anisotropy, the rising and falling antiferromagnetic resonance
branches have different frequencies at zero field. We observe this difference of $d$ and $g$ modes at $H=0$ in \CsCoBr which have frequencies of 95~GHz and 160~GHz,
respectively. SWT \cite{Povarov2} and BOT reproduce this splitting which is a result of chain-alternating easy-plane anisotropy in the model.

To conclude, from the whole variety  of modes of the stripe collinear phase only weak modes $a$, $b$, and $u$ remain unexplained within the proposed theory. Then, the uniform
dynamics (at $\bf k=0$) of $\rm Cs_2CoBr_4$ is not limited by in-chain excitations. However, the quantitative agreement between our theoretical and experimental findings is not
satisfactory above $\approx150~{\rm GHz}\approx0.6$~meV (see Figs.~\ref{dsf1212} and \ref{FvsHb0p5K}) due to pronounced in-chain fluctuations which cannot be properly taken into
account within the first order in $1/n$.

\subsection{UUD phase}

We consider elementary excitations in the UUD phase within the three-spin variant of the BOT as in Ref.~\cite{itrih}, where triangular-lattice Heisenberg antiferromagnet in
magnetic field was discussed. Results obtained are depicted in Fig.~\ref{FvsHb0p5K}. It is seen that theoretically found $T_3$ quasiparticle carrying spin 1 describes well the
most bright lower mode $l$ while we do not observe theoretically its three low-intense satellites ($k$, $m$, and $n$ modes). The splitting of $l$ mode and the appearance of
satellites might be again due to small spin interactions not taken into account (for instance, small inter-plane interaction is known to produce a mode splitting in
triangular-lattice antiferromagnet RbFe(MoO$_4$)$_2$ \cite{RbFeESR}). The spectrum of the lower spin-0 excitation $L_3$ obtained in the BOT does not practically depend on $H$ so
that it is very difficult to observe it using the ESR technique. Such spin-0 quasiparticle lying below magnon modes was found also in Ref.~\cite{itrih} in the UUD state of the
triangular-lattice Heisenberg antiferromagnet.

The mode $o$ obtained experimentally and shown in Fig.~\ref{FvsHb0p5K} is described good (with 15\% accuracy) by $T_4$ excitation found theoretically. Similar to the stripe
phase, there is no good quantitative agreement between our theoretical and experimental findings above $\approx150~{\rm GHz}$ (see Fig.~\ref{FvsHb0p5K}).

Notice that the observed field behavior of the low-lying spin-1 modes $l$ and $o$ ($T_3$ and $T_4$) reproduces qualitatively the field evolution of two lower magnons in
triangular-lattice Heisenberg antiferromagnet in the UUD phase \cite{ChubukovGolosov}: one branch falls to zero at the right boundary of the UUD phase while another branch rises
monotonically from the left limit of the UUD range.

\subsection{Saturated phase}

The saturated phase in which all pseudospins are oriented along the magnetic field has been considered using the four-spin variant of the BOT with the plaquette-like unit cell.
It is seen from Fig.~\ref{FvsHb0p5K} that spectra of lower spin-1 quasiparticles (denoted as $T_6$ and $T_7$) describe reasonably good $w$ and $p$ modes observed experimentally.
We obtain also a high-energy bright spin-1 excitation (corresponding to experimentally found mode $s$ in Fig.~\ref{FvsHb0p5K}) and much weaker spin-2 quasiparticles around it
which correspond presumably to experimentally found modes $t$ and $q$ in Fig.~\ref{FvsHb0p5K}. However, our theory overestimates energies of $s$, $t$, and $q$ modes by more than
40\% (so that we do not show corresponding theoretical lines in Fig.~\ref{FvsHb0p5K}). We attribute this discrepancy, in particular, to the need to take into account the omitted
|$\pm$$3/2\rangle$ one-ion states which come into play at energies $\agt250$~GHz.

Interestingly, there is also another low-lying spin-2 excitation (which can appear in common methods as a two-magnon bound state) whose energy, however, is renormalized downward
considerably by the first $1/n$ corrections. This is an indication that one needs to go beyond the first order in $1/n$ to obtain this spectrum accurately (that is out of the
scope of the present work). We do not show the spectrum of this spin-2 quasiparticle in Fig.~\ref{FvsHb0p5K}.

\subsection{SDW and paramagnet phase}

Theoretical interpretation of ESR data in the SDW and in the paramagnet phases is out of the scope of the present paper.

\section{Summary and conclusion}
\label{conc}

To summarize, we discuss ground-state spin dynamics of strongly anisotropic distorted-triangular-lattice antiferromagnet \CsCoBr by ESR measurements up to 250~GHz and
theoretical analysis. In this compound, interplay of frustration, quantum fluctuations, low dimensionality, and strong anisotropy gives rise to at least six phases in different
orientations of magnetic field $\bf H$ \cite{Povarov1,Povarov2}. Among them are states which are characteristic to 3D antiferromagnets  (collinear stripe state), quasi-1D
magnets (collinear incommensurate spin-density-wave (SDW) phase), and triangular-lattice antiferromagnets (up-up-down (UUD) state with 1/3-magnetization plateau). We observe by
the ESR a rich zoo of excitations in phases arising in $\bf H$ directed along and perpendicular to the magnetic easy axis $b$ which are collected in Figs.~\ref{FvsHb0p5K} and
\ref{FvsHc0p5K}.

Due to strong single-ion anisotropy in spin-$\frac32$ Co$^{2+}$ ions, the low-energy spin dynamics of \CsCoBr which is modeled by Hamiltonian \eqref{ham0} can be described by
strongly anisotropic pseudospin-$\frac12$ Hamiltonian \eqref{hamps} \cite{Povarov1,Povarov2}. Owing to spatial anisotropy of exchange couplings, models \eqref{ham0} and
\eqref{hamps} can be also viewed as spin chains passing along $b$ axis. There are two types of chains in which easy-plane anisotropies are almost perpendicular to each other and
they intersect along $b$ axis (see Fig.~\ref{fig1}). This makes $b$ axis to be easy direction of the whole system. We consider model \eqref{hamps} theoretically using the bond
operator technique (BOT) which is similar in spirit to the well-known spin-wave theory (SWT). However the BOT is more convenient in discussion of "complex" excitations which
appear in other approaches as bound states of ordinary quasiparticles (magnons or triplons) \cite{ibot,iboth,itri,itrih,itrij1j2}.

In the low-field collinear two-sublattice stripe state at ${\bf H}\|b$, the number of modes which we observe using the ESR exceeds essentially the set of magnetic resonance
modes of a conventional  antiferromagnet (see Fig.~\ref{FvsHb0p5K}). However many of these excitations are described using the BOT with model parameters \eqref{param}. Spectra
of spin-1 (magnons) and spin-0 (two-magnon bound states) quasiparticles obtained theoretically are also shown in Fig.~\ref{FvsHb0p5K}. We demonstrate that $d$ and $e$ ESR
signals are described by spin-1 quasiparticles denoted in Fig.~\ref{FvsHb0p5K} as $T_1$ and $T_2$ which are conventional magnons expected in a two-axes quasi-2D magnet strongly
renormalized by quantum fluctuations. In the semi-classical SWT, each of  $T_1$ and $T_2$ are doubly degenerate due to four different spins in the magnetic unit cell (see
Fig.~\ref{fig1}). However quantum fluctuations taken into account more accurately within the BOT lift this degeneracy and produce bright lines $T_1$ and $T_2$ with weak
satellites $T_{1s}$ and $T_{2s}$. The latter is not detected in experiment presumably due to its weakness while the brightest mode $d$ has three satellites $a$, $b$, and $c$
instead of one $T_{1s}$. The increasing number of satellites may be related to small spin interactions which are not included in the model and which frequently lead to modes
splitting in other systems (inter-plane interaction, dipolar forces, etc.). Spin-0 excitations $L_1$ and $L_2$ in Fig.~\ref{FvsHb0p5K} are bound states of two magnons
propagating inside the same and in neighboring spin chains, respectively. It is seen from Fig.~\ref{FvsHb0p5K} that their spectra follow ESR modes $e$ and $f$. The weak
satellite $L_{1s}$ predicted by the theory is not detected experimentally.

It is seen from Fig.~\ref{FvsHb0p5K} that the agreement between the theory and experiment becomes worse above 150~GHz. Using two unit cells in the BOT differing in their
geometry, we argue that this discrepancy is due to profound 1D spin fluctuations inside spin chains which cannot be taken into account properly in the first order in $1/n$ of
the BOT. The confirmation of this comes also from previous neutron scattering and terahertz spectroscopy experiments \cite{Povarov3} in which up to nine sharp anomalies were
observed at the BZ zone center which resemble a Zeeman ladder (ZL) in Izing-like spin chains. These anomalies were related to bound states of two domain walls inside individual
chain confined by inter-plane interaction \cite{Povarov3}. Energies of all these anomalies were described by negative roots of the Airy function (as it is done in conventional
ZLs \cite{mccoy,Rutkevich,Coldea2,Grenier,kimura,sr2,Bera}) except for three lower ones \cite{Povarov3}. We show in Figs.~\ref{dsf1212} and \ref{dsfall} that two low-energy
peaks are reproduced quantitatively within the BOT in model \eqref{hamps} in which the inter-plane interaction is ignored. Besides, these anomalies correspond to spin-1 and
spin-0 quasiparticles $T_1$ and $L_1$ which cannot be parts of the same ZL. And, last but not least, the field evolution of modes $d$, $e$, $f$, and $g$ shown in
Fig.~\ref{FvsHb0p5K} has nothing to do with field behavior of levels in spin-0 and/or spin-1 ZLs (see, e.g., Refs.~\cite{kimura,sr2} and Sec.~\ref{intro}).

Taking into account, however, the successful description in Ref.~\cite{Povarov3} of levels lying above 200~GHz (0.8~meV) within the ZL formalism and bearing in mind our own
results described above, we can state that \CsCoBr combines features in its spin dynamics characteristic to both 2D and 1D anisotropic systems at energies smaller and larger
than 0.8~meV, respectively.

In the UUD state of $\rm Cs_2CoBr_4$, similar to the UUD phase in the Heisenberg triangular-lattice antiferromagnet \cite{itrih}, we observe theoretically two magnon branches
denoted in Fig.~\ref{FvsHb0p5K} as $T_{4,5}$ which go up upon the field increasing and magnon $T_3$ whose energy falls down. Besides, two spin-0 quasiparticles $L_{3,4}$ are
detected one of which lies below magnons that is also in a qualitative agreement with the Heisenberg model \cite{itrih}. It is seen from Fig.~\ref{FvsHb0p5K} that the brightest
ESR modes $l$ and $o$ follows spectra of $T_3$ and $T_4$ magnons. As in the stripe phase, we notice that the most intensive mode $l$ has three weak satellites $k$, $m$, and $n$
whose origin is not clear now and that the agreement between the theory and experiment becomes worse upon the energy increasing. The low-intensive spin-0 modes are not detected
experimentally (in the case of $L_3$, this may be related also to its weak field dependence).

In the saturated state, three high-energy ESR modes can be described only qualitatively using the BOT. We identify the most intensive mode $s$ as a spin-1 excitation which is
surrounded by two spin-2 quasiparticles $t$ and $q$ (corresponding theoretical curves are not shown in Fig.~\ref{FvsHb0p5K}). Lower ESR signals $p$ and $w$ are described
reasonably good by two spin-1 excitations marked in Fig.~\ref{FvsHb0p5K} as $T_6$ and $T_7$.

We leave for the future the theoretical description of two lines of ESR signals in the SDW (Fig.~\ref{FvsHb0p5K}), numerous modes observed in the paramagnet phases
(Fig.~\ref{FvsHb0p5K}), and spin dynamics at ${\bf H}\| c$ (Fig.~\ref{FvsHc0p5K}).

\begin{acknowledgments}

We are grateful to K.Yu.~Povarov and A.~Zheludev for presenting samples and discussion, and to V.N.~Glazkov for discussions. Theoretical study of A.V.~Syromyatnikov is supported
by the Russian Science Foundation (Grant No.\ 22-22-00028). The work on ESR experiment in Kapitza Institute was supported by Russian Science Foundation (Grant No.\ 22-12-00259)

\end{acknowledgments}

\bibliography{lit5CCB}

\end{document}